\documentclass{article}
\usepackage{spconf,amsmath,graphicx,subfigure}
\usepackage{url,cite} 
\usepackage{amssymb}
\usepackage{algorithm}
\usepackage{algorithmic}
\usepackage{multirow}
\usepackage{enumitem}

\title{Revisiting the Problem of Audio-based Hit Song Prediction using Convolutional Neural Networks}

\name{Li-Chia Yang$^*$, Szu-Yu Chou$^*$, Jen-Yu Liu$^*$, Yi-Hsuan Yang$^*$, Yi-An Chen$^\dag$\thanks{This work was partially supported by the Ministry of Science and Technology of Taiwan under Contracts 104-2221-E-001-029-MY3 and 105-2221-E-001-019-MY2.}}
\address{$^*$Research Center for Information Technology Innovation, Academia Sinica, Taiwan\\
         $^\dag$Machine Learning Research Team, KKBOX Inc., Taiwan}
\begin{document}

\maketitle

\begin{abstract}
Being able to predict whether a song can be a hit has important applications in the music industry. Although it is true that the popularity of a song can be greatly affected by external factors such as social and commercial influences, to which degree audio features computed from musical signals (whom we regard as internal factors) can predict song popularity is an interesting research question on its own. Motivated by the recent success of deep learning techniques, we attempt to extend previous work on hit song prediction by jointly learning the audio features and prediction models using deep learning. Specifically, we experiment with a convolutional neural network model that takes the primitive mel-spectrogram as the input for feature learning, a more advanced JYnet model that uses an external song dataset for supervised pre-training and auto-tagging, and the combination of these two models. We also consider the inception model to characterize audio information in different scales. Our experiments suggest that deep structures are indeed more accurate than shallow structures in predicting the popularity of either Chinese or Western Pop songs in Taiwan. We also use the tags predicted by JYnet to gain insights into the result of different models.
\end{abstract}


\section{Introduction}

The popularity of a song can be measured \emph{a posteriori} according to statistics such as the number of digital downloads, playcounts, listeners, or whether the song has been listed in the Billboard Chart once or multiple times. However, for music producers and artists, it would be more interesting if song popularity can be predicted \emph{a priori} before the song is actually released.
For music streaming service providers, an automatic function to identify emerging trends or to discover potentially interesting but not-yet-popular artists is desirable to address the so-called ``long tail'' of music listening \cite{Silk}.
In academia, researchers are also interested in understanding the factors that make a song popular \cite{McClary1991,Lopes1992}. This can be formulated as a pattern recognition problem, where the task is to generalize observed association between song popularity measurements and feature representation characterizing the songs in the training data to unseen songs \cite{Dhanaraj2005}.

Our literature survey shows that this \emph{automatic hit song prediction} task has been approached using mainly two different information sources: 1) \emph{internal factors} directly relating to the content of the songs, including different aspects of audio properties, song lyrics, and the artists; 2) \emph{external factors} encompassing social and commercial influences (e.g. concurrent social events, promotions or album cover design).

The majority of previous work on the internal factors of song popularity are concerned with the audio properties of music.  The early work of Dhanaraj and Logan  \cite{Dhanaraj2005} used support vector machine to classify whether a song will appear in music charts based on latent topic features computed from audio Mel-frequency cepstral coefficients (MFCC) and song lyrics.
Following this work, Pachet \emph{et al.} \cite{Pachet2008} employed a large number of audio features commonly used in music information retrieval (MIR) research and concluded that the features they used are not informative enough to predict hits, claiming that hit song science is not yet a science. Ni \emph{et al.} \cite{Ni2011} took a more optimistic stand, showing that certain audio features such as tempo, duration, loudness and harmonic simplicity correlate well with the evolution of musical trends. However, their work analyzes the evolution of hit songs \cite{MacCallum2012,Mauch2015,Serra2012}, rather than discriminates hits from non-hits.
Fan \emph{et al.} \cite{Fan2013} performed audio-based hit song prediction of music charts in mainland China and UK and found that Chinese hit song prediction is more accurate than the UK version. 
Purely lyric-based hit song prediction was relatively unexplored, except for the work presented by Singhi and Brown \cite{singhi2014hit}.

On the other hand, on external factors, Salganik \emph{et al.} \cite{Salganik2006} showed that the song itself has relatively minor role than the social influences for deciding whether a song can be a hit.  Zangerla \emph{et al.} \cite{Zangerle} used Twitter posts to predict future charts and found that Twitter posts are helpful when the music charts of the recent past are available.

To our best knowledge, despite its recent success in various pattern recognition problems, deep learning techniques have not be employed for hit song prediction.
In particular, in speech and music signal processing, convolutional neural network (CNN) models have exhibited remarkable strength in learning task-specific audio features directly from data, outperforming models based on hand-crafted audio features in many prediction tasks \cite{Choi2016,Ossama,dieleman2014}.

We are therefore motivated to extend previous work on audio-based hit song prediction by using state-of-the-art CNN-based models, using either the primitive, low-level mel-spectrogram directly as the input for feature learning, or a more advanced setting \cite{Liu2016} that exploits an external music auto-tagging dataset \cite{law09magna} for extracting high-level audio features.
Moreover, instead of using music charts, we use a collection of user listening data from KKBOX Inc., a leading music streaming service provider in East Asia. We formulate hit song prediction as a regression problem and test how we can predict the popularity of Chinese and Western Pop music among Taiwanese KKBOX users, whose mother tongue is Mandarin.
Therefore, in addition to testing whether deep models outperform shallow models in hit song prediction, we also investigate how the culture origin of songs affects the performance of different CNN models.

\section{Dataset}
\label{sec:data}

Because we are interested in discriminating hits and non-hits, we find it informative to use the playcounts a song receives over a period of time from streaming services to define song popularity and formulate a regression problem to predict song popularity.
In collaboration with KKBOX Inc., we obtain a subset of user listening records contributed by Taiwanese listeners over a time horizon of one year, from Oct. 2012 to Sep. 2013, involving the playcounts of close to 30K users for around 125K songs.
Based on the language metadata provided by KKBOX, we compile a \emph{Mandarin} subset featuring Chinese Pop songs and a \emph{Western} subset comprising of songs sung mainly in English. There are more songs in the Western subset but the Mandarin songs receive more playcounts on average, for Mandarin is the mother tongue of Taiwanese.

\begin{figure}
\centering
\subfigure{\includegraphics[width=.500\textwidth]{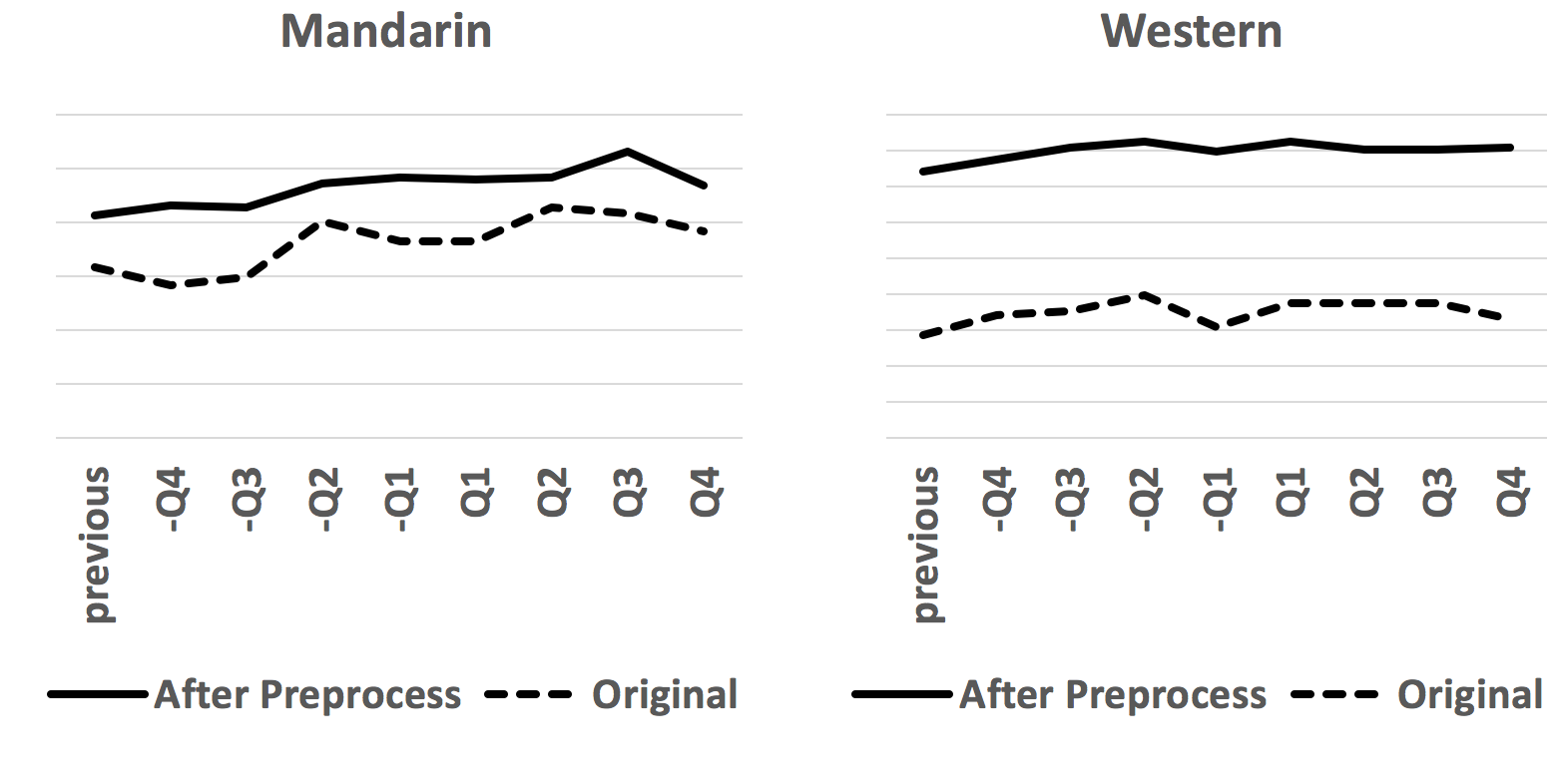}}
\caption{The average playcounts (in log scale) of songs released in different time periods.}
\label{fig:time}
\vspace{-2mm}
\end{figure}

The following steps are taken to gain insights into the data and for data pre-processing. First, as the songs in our dataset are released in different times, we need to check whether we have to compensate for this bias, for intuitively songs released earlier can solicit more playcounts. We plot in Fig. \ref{fig:time} the average playcounts of songs released in different time periods, where Q1 denotes the first three months starting from Oct. 2012 and --Q1 the most recent three months before Oct. 2012, etc. The y-axis is in log scale but the actual values are obscured due to a confidentiality agreement with KKBOX. From the dash lines we see that the average playcounts from different time periods seem to be within a moderate range in the log scale for both subsets, exempting the need to compensate for the time bias by further operations.

\begin{figure}
\centering
\subfigure{\label{fig:a}\includegraphics[width=40mm]{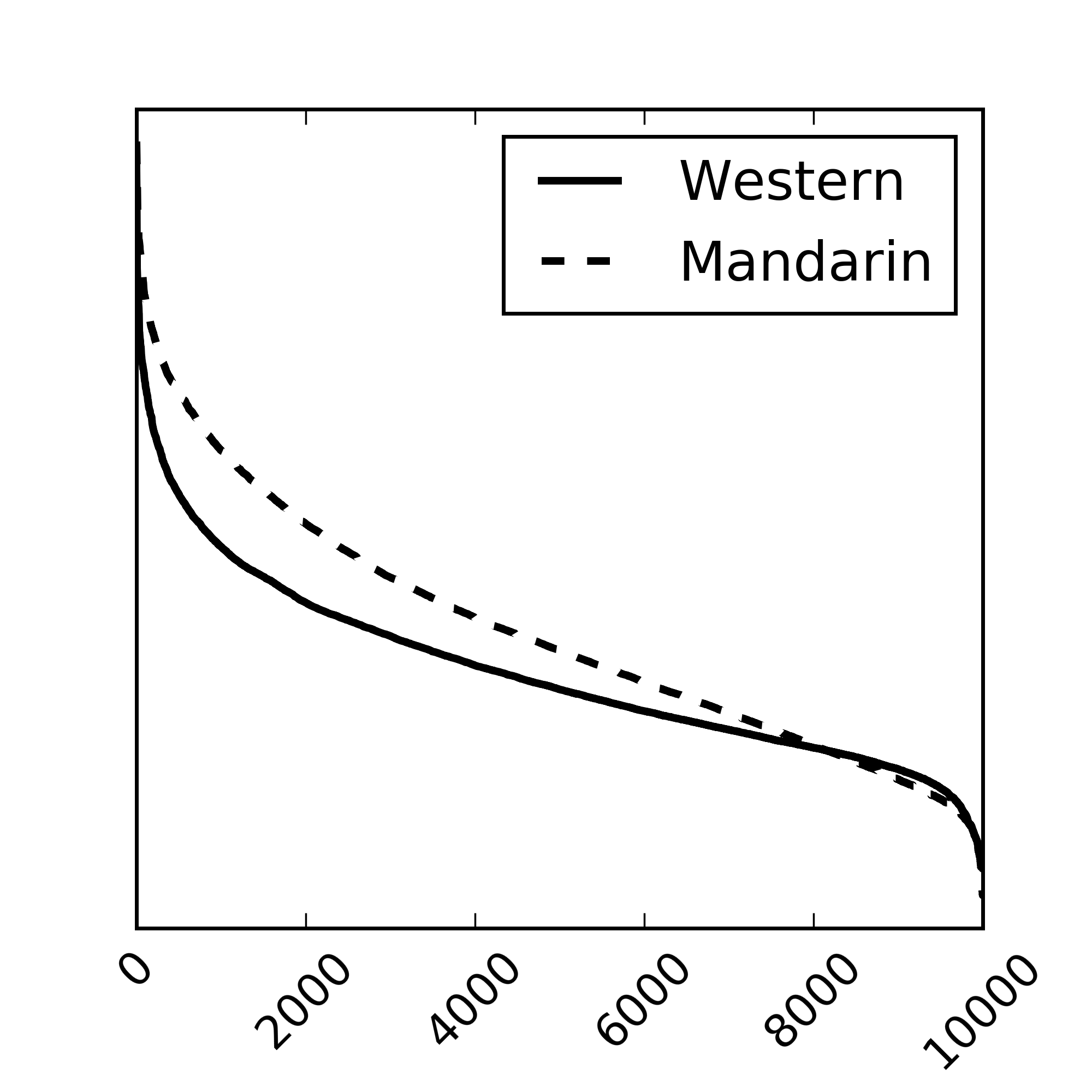}}
\subfigure{\label{fig:b}\includegraphics[width=40mm]{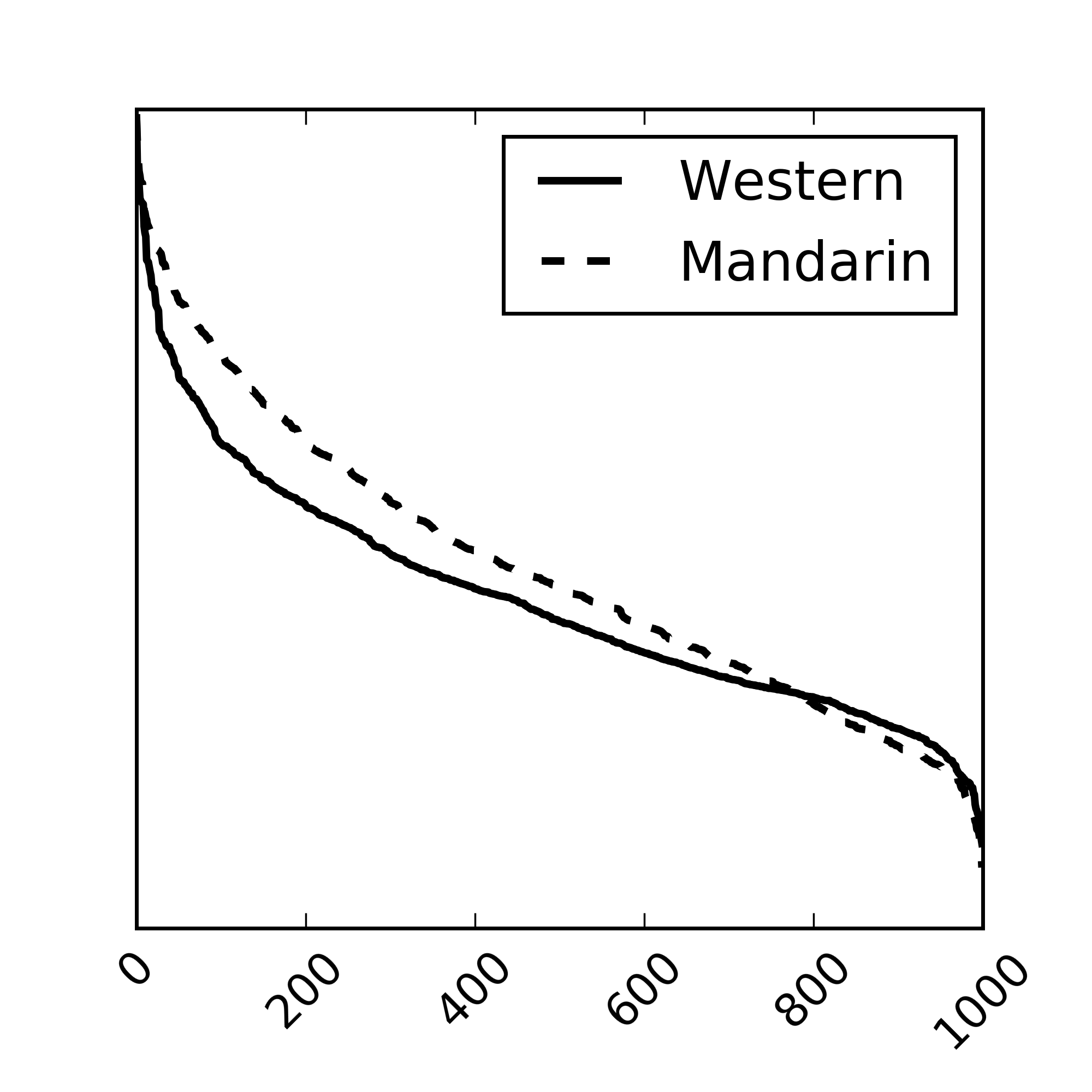}}
\caption[]{The distribution of hit scores (see Section 2 for definition) in the (left) whole and (right) test sets.}
\label{fig:distribution}
\vspace{-2mm}
\end{figure}

Second, we define the \emph{hit score} of a song according to the multiplication of its playcount in log scale and the number of users (also in log scale) who have listened to the song. We opt for not using the playcounts only to measure song popularity because it is possible that the playcount of a song is contributed by only a very small number of users.

Third, to make our experimental results on the two subsets comparable, we sample the same amount of 10K songs in our experiment for both subsets. These songs are those with the highest playcounts within the subset. It can be seen from Fig. \ref{fig:distribution} that the distributions of hit scores of the sampled songs are similar. The solid lines in Fig. \ref{fig:time} show that after this sampling the time bias among the sampled songs remains moderate.

Finally, we randomly split the songs to have 8K, 1K, and 1K songs as the training, validation, and test data for each of the subsets.
Although it may be more interesting to split the songs according to their release dates so as to `learn from the past and predict the future,' we leave this as a future work. Our focus here is to study whether deep models perform better than shallow models in audio-based hit song prediction.

The scale and the time span of the dataset are deemed appropriate for this study.
Unlike previous work on musical trend analysis that may involve more than ten years' worth of data (e.g.\cite{Ni2011}, \cite{Kinoshita2014}),
for the purpose of our work we want to avoid changes in public music tastes and therefore it is better to use listening records collected within a year.

\section{Methods}
\label{sec:method}

We formulate hit song prediction as a regression problem and train either shallow or deep neural network models for predicting the hit scores.
Given the audio representation $\mathbf{x}_n$ for each song $n$ in the training set, the objective is to  optimize the parameters $\Theta$ of our model $f(\cdot)$ by minimizing the squared error between the ground truth $y_{n}$ and our estimate, expressed as $\min_{\Theta }\sum_{n}\left \| y_{n} - f_{\Theta}(\mathbf{x}_{n}) \right \|^2_2$.
As described below, a total number of six methods are considered,
All of them are implemented based on the lasagne library \cite{lasagne}, and the model settings such as learning rate update strategy, dropout rate, and numbers of feature maps per layer are empirically tuned by using the validation set.


\subsection{Method 1 (m1): LR}
As the simplest method, we compute 128-bin log-scaled mel-spectrograms \cite{Dieleman2013} from the audio signals and take the mean and standard deviation over time, leading to a 256-dim feature vector per song. The feature vectors are used as the input to a single-layer shallow neural network model, which is effectively a linear regression (LR) model. The mel-spectrograms are computed by short-time Fourier transform with 4,096-sample, half-overlapping Hanning windows, from the middle 60-second segment of each song, which is sampled at 22 kHz. 
In lasagne, we can implement the LR model by a dense layer. 

\subsection{Method 2 (m2): CNN}
Going deeper, we use the mel-spectrograms directly as the input, which is a 128 by 646 matrix for there are 646 frames per song, to a CNN model. 
Our CNN model consists of two early convolutional layers, with respectively 128-by-4 and 1-by-4 convolutional kernels, and three late convolutional layers, which all has 1-by-1 convolutional kernels.
Unlike usual CNN models, we do not use fully connected layers in the latter half of our model for such \emph{fully convolutional} model has been shown more effective for music \cite{Long2015,Choi2016,Liu2016}.

\subsection{Method 3 (m3): inception CNN}
\begin{figure}
\centering
\subfigure{\includegraphics[width=.500\textwidth]{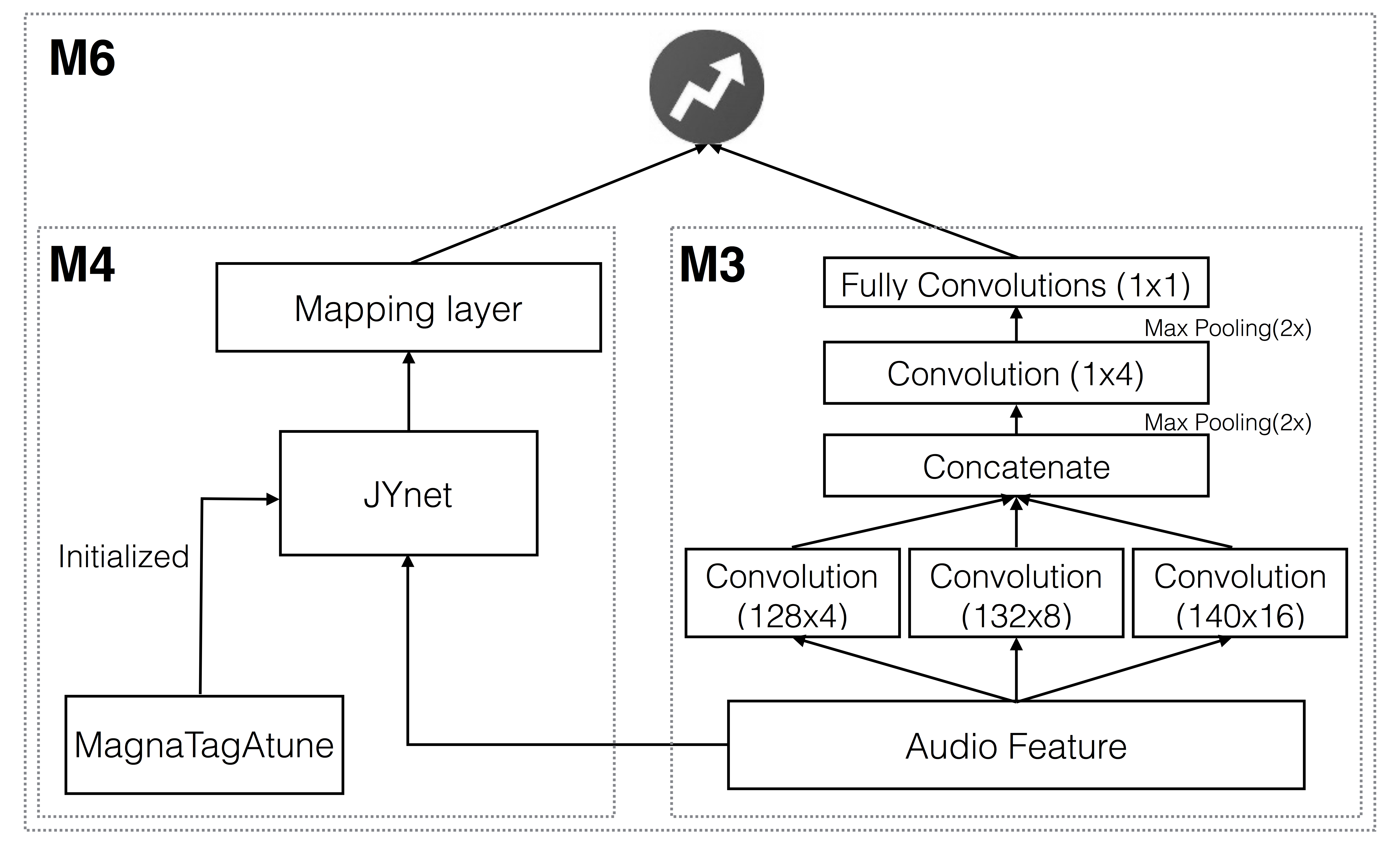}}
\caption[]{Architecture of the investigated CNN models.}
\label{fig: Architecture}
\vspace{-2mm}
\end{figure}

The idea of \emph{inception} was introduced in GoogLeNet for visual problems \cite{Szegedy2014}.
It uses multi-scale kernels to learn features.
We make an audio version of it by adding two more parallel early convolutional layers with different sizes: 132-by-8 and 140-by-16, as illustrated in the bottom-right corner of Fig. \ref{fig: Architecture}. To combine the output of these three kernels by concatenation, the input mel-spectrogram needs to be zero-padded.

\subsection{Method 4 (m4): JYnet (a CNN model) + LR}

While generic audio features such as mel-spectrogram may be too primitive to predict hits, we employ a state-of-art music auto-tagging system referred to as the JYnet \cite{Liu2016} to compute high-level tag-based features.
JYnet is another CNN model that also takes the 128-bin log-scaled mel-spectrograms as the input, but the model is trained
to make tag prediction using the MagnaTagATune dataset \cite{law09magna}.
The output is the activation scores of 50 music tags, including genres, instruments, and other performing related tags such as male vocal, female vocal, fast and slow.
From the output of JYnet (i.e. 50-dim tag-based features), we learn another LR model for predicting hit scores, as illustrated in the bottom-left corner of Fig. \ref{fig: Architecture}.

\subsection{Methods 5 and 6 (m5 \& m6): Joint Training}
We also try to combine (m4) with (m2) or (m3) to exploit information in both the mel-spectrograms and tags, leading to (m5) and (m6).
Instead of simply combining the results of the two models $f_{\Theta_1}(\cdot)$ and $ f_{\Theta_2}(\cdot)$ being combined, we add another layer on top of them for joint training, as illustrated in Fig. \ref{fig: Architecture}.
The learning objective becomes:
\begin{equation}
\min_{w,\Theta_1,\Theta_2}\sum _n{\left \| y_{n} - w f_{\Theta_1}(\mathbf{x}_{n}) - (1-w) f_{\Theta_2}(\mathbf{x}_{n}) \right \|^2_2}\,,
\label{eq:W&D}
\end{equation}
where $w$ determines their relative weight. In this way, we can optimize the model parameters of both models jointly. However, when method 4 is used in joint training we only update the parameters of its LR part, as JYnet is treated as an external, pre-trained model in our implementation.


\section{Experimental Results}
\label{sec:result}

\begin{table*}[!ht]
\centering
\caption{Accuracy of Hit Song Prediction}
\begin{tabular}{l|cccc|cccc}
\hline
\multirow{2}{*}{Method} & \multicolumn{4}{c|}{Mandarin subset} & \multicolumn{4}{c}{Western subset}\\
\cline{2-9}
 & ~~recall~~   & ~nDCG~  & Kendall   & Spearman & ~~recall~~    & ~nDCG~   & Kendall   & Spearman \\ \hline
\textbf{(m1)} audio+LR      & 0.1900          & 0.1997          & 0.1679          & 0.2480    & 0.1400            & 0.1271            & 0.0674            & 0.1002            \\
\textbf{(m2)} audio+CNN      & 0.2300          & 0.2334          & 0.1806          & 0.2678     & 0.1300            & 0.1294            & 0.1031            & 0.1564            \\
\textbf{(m3)} audio+inception~CNN    & 0.2500          & 0.2369          & 0.2286          & 0.3374     & 0.1800            & 0.1989            & 0.1093            & 0.1636        \\ \hline
\textbf{(m4)} tag+LR      & 0.2400          & 0.2372          & 0.1671          & 0.2473     & 0.2000             & 0.1774            & 0.0918            & 0.1372         \\ \hline
\textbf{(m5)} (m2)+(m4)       & 0.2500          & 0.2558          & 0.2018          & 0.2971       & 0.1800            & 0.1791            & 0.1300            & 0.1941         \\
\textbf{(m6)} (m3)+(m4)     & {0.3000} & {0.2927} & {0.2665} & {0.3894} & {0.2100} & {0.2413} & {0.1341} & {0.1996} \\ \hline
\end{tabular}
\label{table:result}
\end{table*}


We train and evaluate the two data subsets separately. For evaluation, the following four metrics are considered:
\begin{itemize}[noitemsep,topsep=3pt,parsep=0.5pt,partopsep=0.5pt]
\item Recall@100: Treating the 100 songs (i.e. 10\%) with the highest hit scores among the 1,000 test songs as the hit songs, we rank all the test songs in descending order of the predicted hit scores and count the number of hit songs that occur in the top 100 of the resulting ranking.
\item nDCG@100: normalized discounted cumulative gain (nDCG) is another popular measure used in ranking problems \cite{Wang2013}. It is computed in a way similar to recall@100, but the positions of recalled hit songs in the ranking list are taken into account.
\item Kendall's $\tau$: we directly compare the ground truth and predicted rankings of the test songs in hit scores (without defining which songs are hit songs) and compute a value that is based on the number of correctly and incorrectly ranked pairs \cite{Kendall1990}.
\item Spearman's $\rho$: the rank correlation coefficient (considering the relative rankings but not the actual hit scores) between the ground truth and predicted rankings.
\end{itemize}

The result is shown in Table \ref{table:result}, which is obtained by averaging the result of 10 repetition of each method. The following observations can be made.
First, by comparing the result of (m1), (m2) and (m3), we see that better result in most of the four metrics is obtained by using deeper and more complicated models for both subsets. This suggests the effectiveness of deep structures for this task. Furthermore, by comparing the result of the two subsets, we see that audio-based hit song prediction is easier for the Mandarin subset, confirming the findings of Fan \emph{et al.} \cite{Fan2013}.

Second, as both (m1) and (m4) use LR for prediction, by comparing their result we see that the tag-based method (m4) outperforms the simple audio-based method (m1) in all the four metrics for the Western subset, demonstrating the effectiveness of the JYnet tags. This is however not the case for the Mandarin subset for Kendall's $\tau$ and Spearman's $\rho$.

Third, from the result of (m5) and (m6), we see that the joint learning structure can further improve the result for both subsets. The best result is obtained by (m6) in all metrics.

\begin{figure}
\centering
\includegraphics[width=.92\columnwidth]{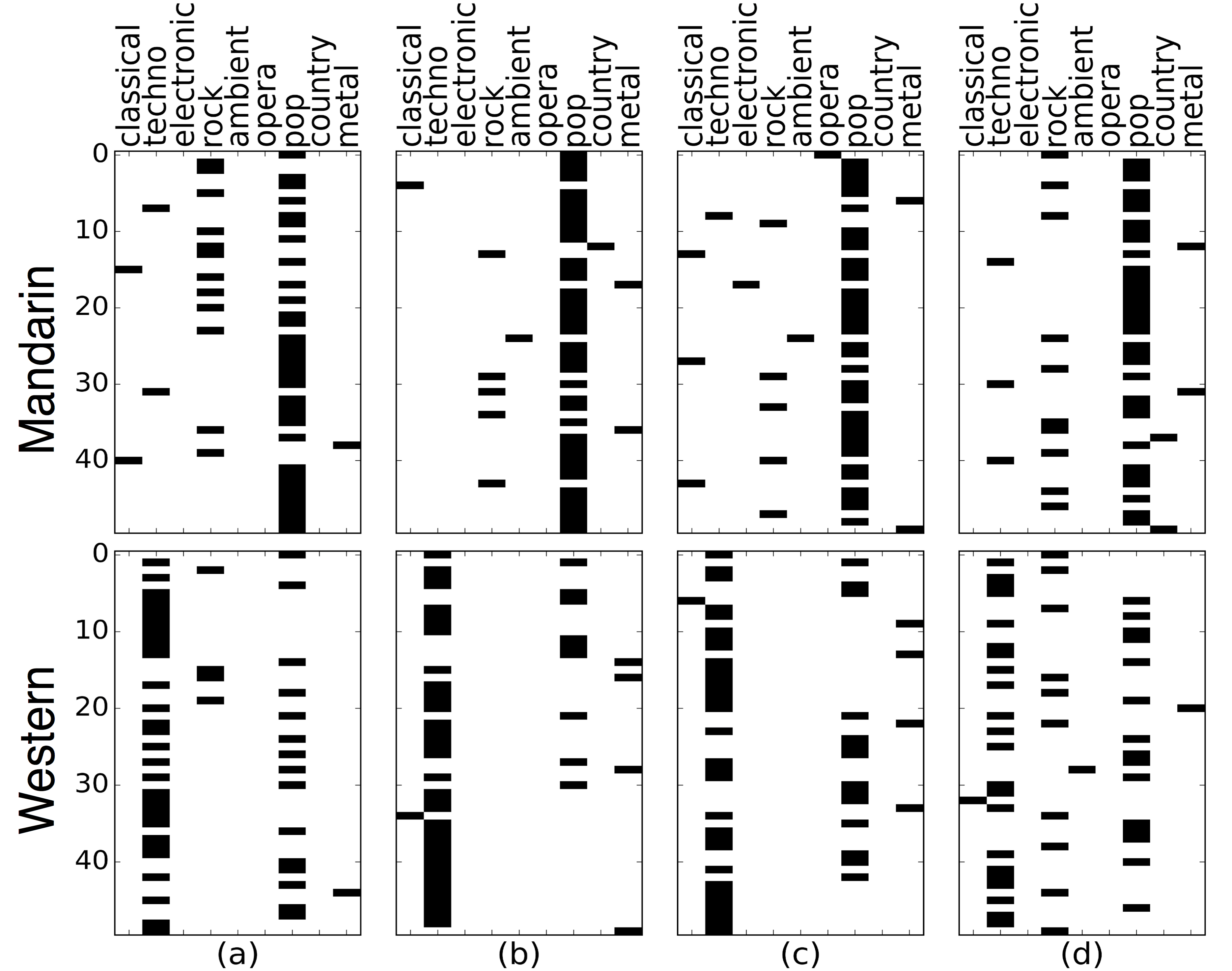}
\caption{The predominate tags (predicted by JYnet) for the top-50 hit songs determined by different methods for the (top) Mandarin and (bottom) Western subsets. From left to right: (a) the tag-based model (m4), (b) the audio-based model (m3), (c) the hybrid model (m6), and (d) the ground truth.}
\label{fig:genre}
\vspace{-2mm}
\end{figure}

To gain insights, we employ JYnet to assign genre labels to all the test songs and examine the distribution of genres in the top-50 hit songs determined by either automatic models or the ground truth.
For each song, we pick the genre label that has the strongest activation as predicted by JYnet. The resulting genre distributions are shown in Fig. \ref{fig:genre}. We see from the result of ground truth that the Western hits have more diverse genres. The predominance of `Pop' songs in the Mandarin subset might explain why 1) hit song prediction in this subset is easier and 2) (m4) alone cannot improve $\tau$ and $\rho$.
Moreover, for the Western subset, we see that the genre distribution of (m4) is more diverse than that of (m3), despite that (m3) achieves slightly higher nDCG and Spearman's $\rho$.
This might imply that the ability to match the genre distribution of the ground truth is another important performance indicator.  

\section{Conclusion}
\label{sec:concl}
In this paper, we have introduced state-of-the-art deep learning techniques to the audio-based hit song prediction problem.
Instead of aiming at classifying hits from non-hits, we formulate it as a regression problem. 
Evaluations on the listening data of Taiwanese users of a streaming company called KKBOX confirms the superiority of deep structures over shallow structures in predicting song popularity.
Deep structures are in particular important for Western songs, as simple shallow models may not capture the rich acoustic and genre diversity exhibited in Western hits.
For future work, we hope to understand what our neural network models actually learn, to compare against more existing methods (preferably using the same datasets), and to investigate whether our models can predict future charts or emerging trends.

\bibliographystyle{IEEEbib2}
\bibliography{POP}

\end{document}